\documentclass[preprint, review, sort&compress, usenames, 3p,times,dvipsnames, 12pt]{elsarticle}

\usepackage{amssymb}

\usepackage{forest}

\definecolor{folderbg}{RGB}{124,166,198}
\definecolor{folderborder}{RGB}{110,144,169}

\def\Size{4pt}
\tikzset{
      folder/.pic={
        \filldraw[draw=folderborder,top color=folderbg!50,bottom color=folderbg]
          (-1.05*\Size,0.2\Size+5pt) rectangle ++(.75*\Size,-0.2\Size-5pt);  
        \filldraw[draw=folderborder,top color=folderbg!50,bottom color=folderbg]
          (-1.15*\Size,-\Size) rectangle (1.15*\Size,\Size);
      }
    }

\usepackage{xcolor}

\usepackage{tikz}
\usepackage[toc,page]{appendix}
\usepackage{longtable}
\usepackage{tabularx}
\usepackage{tabu}
\usepackage{amsmath}
\usepackage{dcolumn}
\usepackage{bm}
\usepackage{mathrsfs}
\usepackage{float}
\usepackage{caption}

\usepackage{subcaption}
\usepackage[linesnumbered,ruled,vlined]{algorithm2e}
\usepackage{booktabs} 
\usetikzlibrary{arrows,decorations.markings,shapes.geometric, arrows}
\usetikzlibrary{positioning}
\usetikzlibrary{calc}
\usetikzlibrary{shapes,backgrounds}
\usepackage{setspace}
\usepackage{amsfonts}
\usepackage{ctable} 
\usepackage{natbib}
\usepackage{placeins}
\definecolor{ashgrey}{rgb}{0.7, 0.75, 0.71}
\tikzoption{base font}{\def\tikz@base@textfont{#1}}
\tikzoption{font}{\def\tikz@textfont{\tikz@base@textfont#1}}
\tikzset{
	base font=\sffamily,
}

\restylefloat{table}
\newcommand{\ei}{\vec{e}_i}
\newcommand{\wi}{w_i}

\newcommand{\vx}{\vec{x}}

\newcommand*{\GridSize}{8}

\newcommand*{\ColorCells}[1]{
\foreach \x/\y/\color in {#1} {
  \node [fill=\color, draw=none, thick, minimum size=1cm]
  at (\x-.5,\GridSize+0.5-\y) {};
}%
}%

\begin{document}

\begin{frontmatter}



\title{LBfoam: An open-source software package for the simulation of foaming using the Lattice Boltzmann Method}


\author[label1]{M. Ataei}
\author[label1]{V. Shaayegan}
\author[label2]{F. Costa}
\author[label3]{S. Han}
\author[label1]{C. B. Park}
\author[label1]{M. Bussmann\corref{cor1}}
\address[label1]{Department of Mechanical Engineering, University of Toronto, 5 King's College Rd, Toronto, ON M5S 3G8, Canada}
\address[label2]{Autodesk, Inc., 259-261 Colchester Rd., Kilsyth, VIC. 3137, Australia}
\address[label3]{Autodesk, Inc. 2353 North Triphammer Rd., Ithaca, NY 14850, USA}
\linespread{2}

\cortext[cor1]{Corresponding author}

\begin{abstract}
This paper presents a 2D/3D Free Surface Lattice Boltzmann Method simulation package called LBfoam for the simulation of foaming processes. The model incorporates the essential physics of foaming phenomena: gas diffusion into nucleated bubbles, bubble dynamics and coalescence, surface tension, the stabilizing disjoining pressure between bubbles, and Newtonian and non-Newtonian rheological models. The software can simulate the growth and interaction of bubbles, and predict final foam structures. The implementation is based on the Palabos library (in C++), which enables large-scale parallel simulations. The software is freely available under the GNU Affero General Public License version 3 at:

https://github.com/mehdiataei/LBfoam

\end{abstract}

\begin{keyword}
 foaming \sep Lattice Boltzmann Method \sep bubble growth \sep numerical model \sep open-source software \sep free surface flow
\end{keyword}

\end{frontmatter}

\section{Introduction}
Foams are used in a wide range of products in the automotive, aerospace, furniture, packaging and insulation industries due to desired mechanical characteristics such as weight reduction, improved strength per weight, and energy absorption. Foams are formed by nucleation and growth of gas bubbles in a liquid medium. Initially either, a metered amount of gas is dissolved in the liquid, or the gas is produced by a foaming agent that is mixed with the liquid. A rapid drop in solubility (e.g.\ induced by a pressure drop) gives rise to nucleated gas bubbles, which grow due to the diffusion of gas molecules from the liquid. Ultimately, bubbles are separated by thin liquid regions (lamellae), forming a cellular structure. Simulating this foaming process is essential for developing foam structures with desired properties as, for example, homogeneous foams tend to improve the mechanical strength per weight of a final structure. 

Developing a numerical model for foam evolution is challenging due to the complexity of the physical phenomena. There have been many attempts in the literature to model the foaming process by introducing simplifications. Various single bubble models (i.e., the ``Cell Model") estimate the growth rate of an isolated bubble as a 1D diffusion problem, but they neglect bubble interactions and assume that the foam is a periodic array of spherical bubbles (e.g.\ \cite{PATEL19802352,Amon1984,Arefmanesh1990, Arefmanesh1991,WANG2019189,Arefmanesh1995}). A comparison of these models can be found in \cite{Elshereef2010}. In a number of 2D models, the transport equations are solved for only one section of a lamella separating adjacent bubbles, by assuming that the bubbles are organized in known arrangements such as hexagonal arrays \cite{Everitt,EVERITT200646,EVERITT200660}. The arbitrary Lagrangian-Eulerian sharp interface algorithm has also been used to simulate bubble growth in 2D \cite{YUE20072229}, although this model is prone to errors for large bubble deformations and cannot capture bubble coalescence and rupture. The Shan-Chen model \cite{shanchen} based on the Lattice Boltzmann Method (LBM) has shown promising results for the simulation of bubble growth and interaction for small liquid-gas density ratios \cite{BARZEGARI20191258}.

There is another class of foaming simulations over much larger length scales, where the interest is not individual bubble growth and interactions, but rather to model foam flow into a given mold geometry, i.e.\ mold-filling. These models omit details of foaming at the bubble-scale, but use bubble scale models to extract certain constants and information that are then incorporated into particle tracers, surrogate models, or population balance equations, to approximate the effects of bubble growth on macroscale fluid properties such as fluid density and viscosity (e.g.\ \cite{Karimi2017, Geier2}).


Here we present a foaming simulation software, which is a modified version of the Free Surface Lattice Boltzmann Method (FSLBM) that is originally developed to model fluid flows when a large fraction of a domain is occupied by gas \cite{THUREY2009221, THURNEY2, Korner2002,Korner2005, LatticeFree2, FS1,FS2,FS4,FS5,FS8}. The pressure and gas content of each bubble is monitored throughout the simulation, and sudden topological variations due to bubble coalesce, splitting, merging, and bursting are accounted for. The numerical model couples the gas advection-diffusion with the fluid flow model, for which bubble interfaces are arbitrarily-shaped immersed boundaries at which gas concentration must be specified. A short-range force between bubble interfaces (i.e.\ disjoining pressure) is considered to allow formation of stable lamella between bubbles.

This simulation model encompasses all of the essential physics of foaming, and can be used to analyze 2D and 3D foaming processes such as polymer foam injection molding and metal foaming for prediction of the final foam structure, and for understanding phenomena such as topological changes (e.g.\ bubble coalescence), the occurrence of foam drainage, and the dissolution of bubbles.

The most closely related work is that of K\"{o}rner et al.\ \cite{Korner2002} on a 2D FSLBM foaming model. They simulated the expansion of gas bubbles in molten aluminum, and studied the influence of viscosity, surface tension, and mold constraints on the final structure of a foam. Since their source code is not publicly available, the scope and capabilities of the model are not known. And although a 3D version of the model has been developed \cite{Korner2005}, advection-diffusion of gas is not included, and bubble growth is modeled by increasing bubble gas content in proportion to bubble surface area.

The software presented here extends the Palabos library \cite{palabos} in a non-intrusive way, so that the user can easily access useful features of Palabos including efficient MPI parallelization for large-scale simulations, access to various collision operators, the use of complex geometries for the simulation domain using STL meshes, different viscosity models such as Newtonian, Smagorinsky, and Carreau-Yasuda, and various boundary conditions.

In the remainder of the paper, we present our formulation of FSLBM for foaming, followed by several validation and simulation results of different foaming phenomena. 

\section{Free Surface Lattice Boltzmann Method}

The Lattice Boltzmann Method (LBM) is widely used to perform CFD simulations. The Chapman–Enskog expansion \cite{Chapman} shows that a discretized Boltzmann equation with a collision kernel ($\hat{C}$) can solve the incompressible Navier-Stokes equations for fluid flow. The discretized lattice Boltzmann equation reads

\begin{equation}
f_i(\vx+\ei \delta_t,t+\delta_t) = f_i(\vx,t) + \hat{C}(\vx,t)
\label{eqn:LBE}
\end{equation}

The particle distribution function (PDF) for each lattice direction at time $t$ is given by $f_i(\vx,t)$ where $\delta_t$. In LBM, space is discretized with square (in 2D) or cubic lattices (in 3D) with a finite set of $N$ discrete lattice velocities $\vec{e}_i$ ($i=0,\dots, N-1$). For example, in 2D and 3D, one can use D2Q9 and D3Q19 lattices, respectively. The velocity sets for the D3Q19 lattice are:

\begin{equation}
\small
\ei^\top=\begin{cases}
(0,0,0), & i=0\\
(\pm1,0,0), (0,\pm1,0), (0,0,\pm1), & i = 1,2,3,4,5,6 \\
(\pm1,\pm1,0), (\pm1,0,\pm1), (0,\pm1,\pm1) & i = 7,\dots,18 \\

\end{cases}
\end{equation}

\begin{figure}[H]
	\centering
	\begin{subfigure}[b]{0.35\textwidth}
		\includegraphics[width=\textwidth]{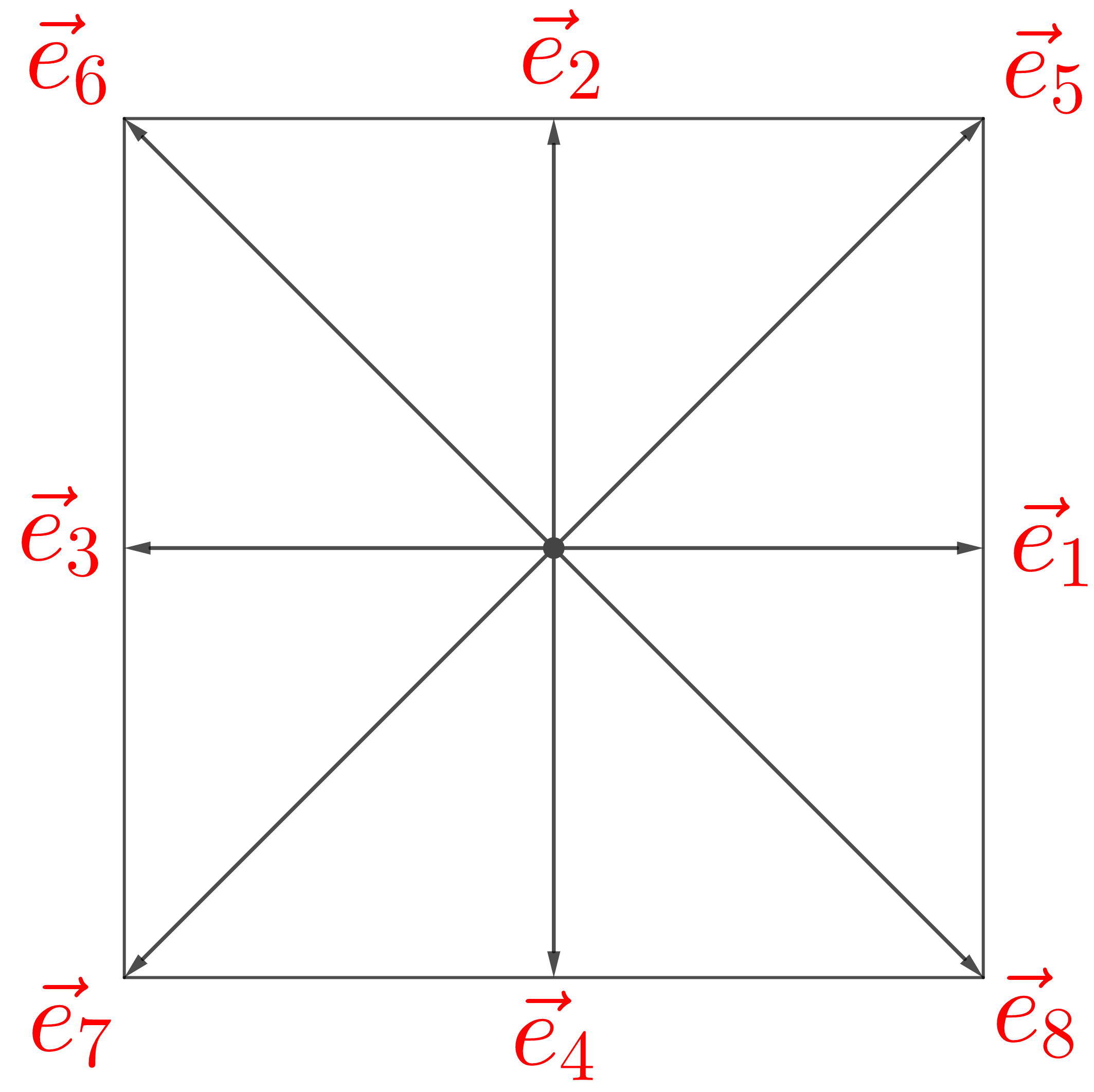}
		\caption{D2Q9 lattice. $w_0=4/9$, $w_1,\dots,w_4=1/9$, $w_5,\dots,w_{9}=1/36$.}
		\label{fig:D2Q9}
	\end{subfigure}  \hspace{5mm}
	\begin{subfigure}[b]{0.35\textwidth}
	\includegraphics[width=\textwidth]{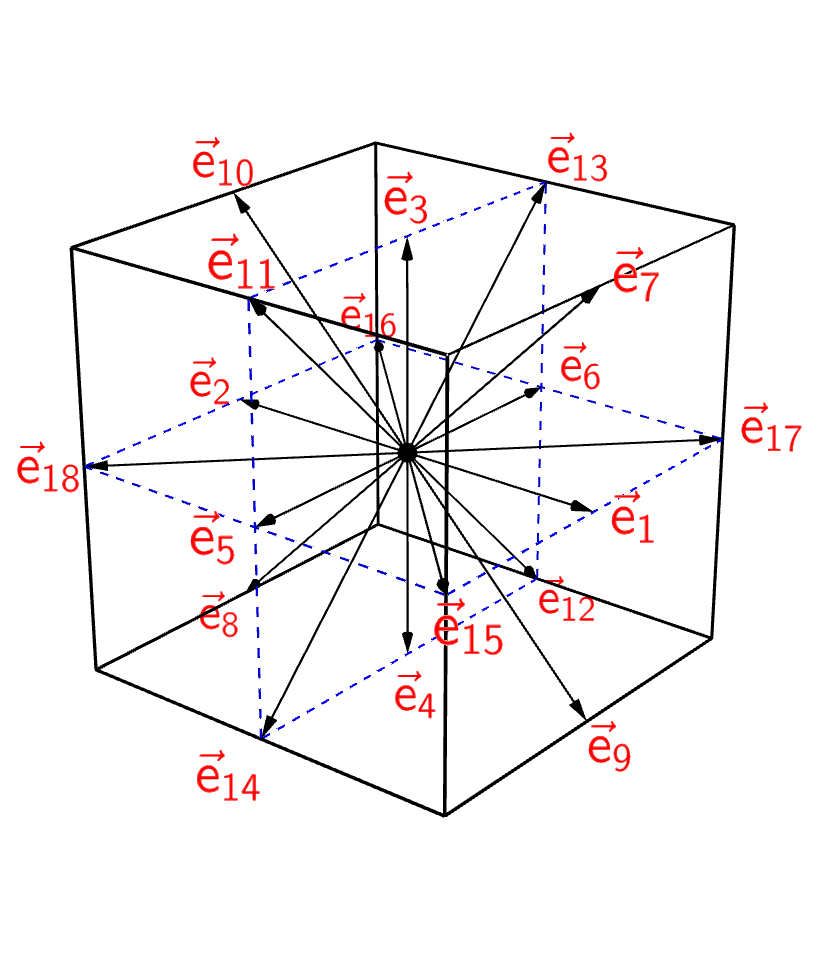}
	\caption{D3Q19 lattice. $w_0=1/3$, $w_1,\dots,w_6=1/18$, $w_7,\dots,w_{18}=1/36$.}
	\label{fig:D3Q19}
	\end{subfigure}
	\caption{2D and 3D lattices that are used to solve the Navier-Stokes equation using LBM.}\label{fig:latticesfluid}
\end{figure}

\noindent Fig.\ \ref{fig:latticesfluid} shows the lattices and their respective weights $\wi$. The spatial resolution of a domain in 3D corresponds to the lattice width $\delta_l$, so that a box of length $l_x$, $l_y$ and $l_z$ is comprised of $\frac{l_x}{\delta_l}$, $\frac{l_y}{\delta_l}$ and $\frac{l_z}{\delta_l}$ lattices (also referred to as \textit{cells} in this paper). 

Equation (\ref{eqn:LBE}) can be decomposed into collision and streaming steps:

\begin{equation}
\text{\textbf{Collision}}: f^c_i(\vx,t) = f_i(\vx,t) + \hat{C}(\vx,t)
\label{eqn:collisionstep}
\end{equation}

\begin{equation}
\text{\textbf{Streaming}}: f_i(\vx+\ei \delta_t,t+\delta_t) = f^c_i(\vx,t)
\label{eqn:streamingstep}
\end{equation}

\noindent where $f^c_i$ denotes the post-collision PDF. The collision operator $\hat{C}$ is often replaced by the Bhatnagar-Gross-Krook (BGK) model for isothermal systems with an external force term $F_i$ \cite{He1997}:

\begin{equation}
\
\hat{C}(\vx,t) = - \frac{\delta_t}{\tau}\left(f_i(\vx,t)-f_i^{eq}(\vx,t)\right) - F_i
\
\label{eqn:collision}
\end{equation}

\noindent where $\tau$ is the time interval between particle collisions. The equilibrium distribution function $f_i^{eq}$ is the truncated second-order Taylor expansion of the Maxwell distribution function with respect to flow velocity $\vec{u}$:

\begin{equation}
f^{eq}_i(\rho,\vec{u}) = \rho \wi \left( 1 + \frac{\ei^\top \vec{u}}{c_s^2} + \frac{(\ei^\top \vec{u})^2}{2c_s^4} - \frac{\vec{u}^2}{2c_s^2} \right)
\label{eqn:equilibrium}
\end{equation}

\noindent $\rho$ is the density and $c_s$ is the speed of sound, that has a value of $1/\sqrt3$ for the D2Q9 and D3Q19 lattices. Both $\rho$ and $\vec{u}$ are functions of $\vx$ and $t$. The force term $F_i$ in Eq.\ \ref{eqn:collision} accounts for external forces such as gravity $g$. For an external force generating an acceleration $\vec{a}$, we have \cite{He1997}:

\begin{equation}
F_i = \rho \wi \left(\frac{\ei^\top \vec{u}}{2c_s^4} - \frac{\vec{u} - \ei}{c_s^2}\right) \cdot \vec{a}
\end{equation}

\noindent Palabos provides access to several other collision operators including the Multiple-Relaxation-Time (MRT), Two-Relaxation-Time (TRT), Regularized Lattice Boltzmann (RLB), and the Entropic Lattice Boltzmann (ELB) models, that can be used instead of the BGK model. A comparative study of these models can be found in \cite{EZZATNESHAN2019158}.

The macroscopic quantities $\rho$ and $\vec{u}$ are calculated by summation of PDFs:

\begin{equation}
\rho(\vx,t) =  \sum_{i=0}^{N} f_i
\end{equation}

\begin{equation}
\vec{u}(\vx,t) = \frac{1}{\rho} \sum_{i=0}^{N} f_i \ei
\label{eqn:velocity}
\end{equation}

\noindent and pressure is proportional to the density:

\begin{equation}
p(\vx,t) = \rho(\vx,t) c_s^2
\end{equation}

\noindent The bounce-back scheme \cite{He1997Bback} is used at wall nodes, meaning that particles colliding with a wall have their momentum reversed, which translates to a no-slip boundary condition. Hereafter, we use $\delta_l=\delta_t=1$ for simplicity.

\subsection{Interface Capturing}

\label{sec:interfacetracking}

Interfaces between gas and liquid phases in a foam are captured using a volume-of-fluid (VOF) method, where a dimensionless scalar variable $\alpha$ is introduced to track the motion of interfaces throughout the computational domain by solving a pure advection transport equation:

\begin{equation}
    \frac{\partial \alpha}{\partial t} + \mathbf{u} \cdot \nabla \alpha = 0
    \label{eqn:vof}
\end{equation}

\noindent $\alpha$ is equal to one in liquid cells, zero in gas cells, and it has a value between zero and one in all interface cells (see Fig.\ \ref{fig:vof}):

\begin{equation}
\alpha=\begin{cases}
0, & \forall \vx \in G\\
1, & \forall \vx \in L\\
0 < \alpha < 1 & \forall \vx \in I
\end{cases}
\end{equation}

\noindent $G$, $L$, and $I$ represent gas, liquid, and interface cells, respectively.

\noindent Here, instead of solving Eq.\ \ref{eqn:vof} explicitly, we use a fast mass tracking algorithm \cite{Korner2005} that is commonly used in FSLBM simulations, which takes advantage of the evolution of PDFs to track $\alpha$ in the domain. In this algorithm, $\alpha$ is defined as:

\begin{equation}
\alpha = \frac{M}{\rho \delta^3_l}
\label{eqn:alpha}
\end{equation}

\noindent where $M$ is the liquid mass in a cell. 

\noindent The value of $\alpha(\vx, t+1)$ is found by calculating the mass exchange between neighboring cells $\vx$ and $\vx + \ei$:

\begin{figure}
	\centering
	\fontfamily{lmtt}\selectfont
	
\begin{tikzpicture}[scale=0.5,every node/.style={scale=0.5}]

\begin{scope}[thick,local bounding box=name]
\ColorCells{3/4/Cyan, 4/4/Cyan, 5/4/Cyan, 6/4/Cyan, 3/5/Cyan, 1/6/Cyan, 2/5/Cyan, 1/7/Cyan, 2/6/Cyan, 6/5/Cyan, 7/5/Cyan, 8/5/Cyan}
\foreach \i in {1,...,8} {
	\ColorCells{\i/8/blue}
}
\foreach \i in {2,...,8} {
	\ColorCells{\i/7/blue}
}
\foreach \i in {3,...,8} {
	\ColorCells{\i/6/blue}
}
\foreach \i in {4,5} {
	\ColorCells{\i/5/blue}
}

\draw (0, 0) grid (\GridSize, \GridSize);
\end{scope}

\draw[very thick] (0,1) .. controls (3,5)  .. (8,3);

\draw[->,thick] (-3,4) -- (3.5,2) node [pos=0.0,xshift=-7ex,yshift=7ex,below] {\huge  liquid cells ($\alpha = 1$)};
\draw[->,thick] (3.5,8.5) -- (3.5,4.5) node [pos=0.0,xshift=18ex,yshift=6ex,below] {\huge interface cells ($0 < \alpha < 1$)};
\draw[->,thick] (-2,8.5) -- (1.5,5.5) node [pos=0.0,xshift=-7ex,yshift=6ex,below] {\huge gas cells($\alpha = 0$)};

\end{tikzpicture} \\
	\caption{Demonstrating the VOF scalar field.}
	\label{fig:vof}
\end{figure}
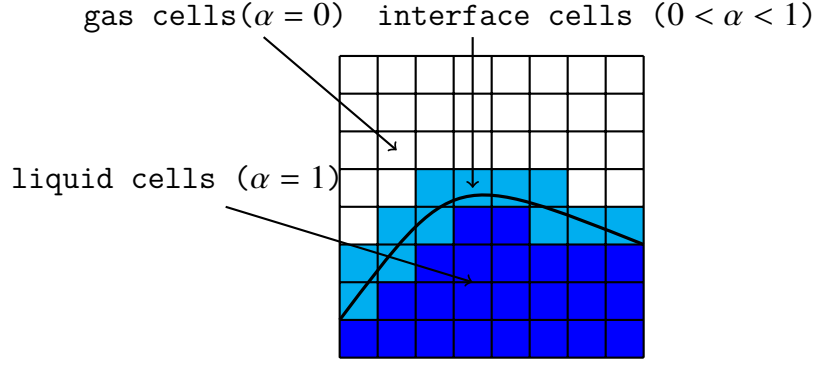

\begin{center}
\begin{equation}
\alpha(\vx,t+1)=\alpha(\vx,t)+\frac{1}{\rho(\vx,t)}\sum_{i=0}^{N} \Theta	(f_{\bar{i}}(\vx+\ei,t) -f_i(\vx,t))
\label{eqn:massexchange}
\end{equation}
\end{center}

\noindent where $\bar{i} = -i$. The parameter $\Theta$ in Eq.\ \ref{eqn:massexchange} weights the mass exchange between two interface cells by their average volume fraction:

\begin{equation}
\Theta	=\begin{cases}
0, & \vx+\ei \in G\\
1, & \vx+\ei \in L\\
\frac{1}{2}\left(\alpha(\vx,t)+\alpha(\vx+\ei,t)\right)& \vx+\ei \in I
\end{cases}
\end{equation}

\noindent Equation (\ref{eqn:massexchange}) conserves mass locally since the mass that leaves one cell is distributed to neighboring cells. An interface cell becomes a liquid cell when $\alpha(\vx,t) \geq 1$, and an interface cell becomes an empty cell when $\alpha(\vx,t) \leq	0$. Where there is an excess or shortage of mass in a cell (i.e.\ $\alpha(\vx,t) > 1$ or $\alpha(\vx,t) < 0$), the surplus or shortage is uniformly distributed to neighboring interface cells. Other interface capturing and tracking methods are also available e.g.\ \cite{MEHRAVARAN2008223, HUGHES1981329}.

\subsection{Free Surface Formulation}

The liquid-gas interaction is modelled as a free surface. Since the ratios of liquid to gas density and viscosity are often very large, the hydrodynamics of the gas phase are ignored for simplicity. Only the influence of the pressure of the gas phase is considered.

The free surface simplification means that the lattice Boltzmann equation is not solved for in the gas cells. As a result, in the streaming step (Eq.\ \ref{eqn:streamingstep}), the PDFs arriving from gas cells are undefined in interface cells. These undefined PDFs ($f_i^u$) are imposed such that the pressure boundary condition at the free surface is enforced, and so that the velocity of gas cells is equal to the adjacent liquid velocity $\vec{u}_I$. For this purpose, the unknown distribution functions at the interface are set as \cite{Korner2005}:

\begin{equation}
f^u_i (\vx - \ei, t+1) = f^{eq}_i(\rho_g,\vec{u}_I) + f^{eq}_{\bar{i}}(\rho_g,\vec{u}_I) - f_{\bar{i}}(\vx -\ei, t + 1)
\label{eqn:reconstruct}
\end{equation}

The effect of surface tension is considered in the model by modifying the gas density $\rho_g$ in Eq.\ \ref{eqn:reconstruct} \cite{Korner2005}:

\begin{equation}
\rho_g = \frac{p_g - 2\gamma\kappa(\vx)}{c^2_s}\qquad \vx \in I
\label{eqn:rhob}
\end{equation}

\noindent $\gamma$ is the surface tension, and $\kappa(\vx)$ is the local curvature. Palabos provides different methods for curvature calculation, including using the gradient of a smoothed $\alpha$ scalar field \cite{CUMMINS2005425}, and a height function \cite{heightFunction3D}. In Eq.\ \ref{eqn:rhob}, gas pressure $p_g$ for each individual bubble $i$ is given by the ideal gas law:

\begin{equation}
\label{eqn:idealgas}
p^i_g = \frac{m^i_g  \mathscr{R} T}{V^i}
\end{equation}

\noindent where $m^i_g$, $\mathscr{R}$, $T$, and $V^i$ are the mass of a bubble, gas constant, temperature, and volume of a bubble.

\subsection{Viscosity}

In the collision operator (Eq.\ \ref{eqn:collision}), $\tau$ is related to kinematic viscosity $\nu$ by \cite{He1997}
\begin{equation}
\nu = c_s^2\left(\tau - \frac{1}{2}\right)
\label{eqn:tau}
\end{equation}

\noindent For a Newtonian fluid, $\tau$ is a constant. For shear thinning polymers, the Carreau-Yasuda model \cite{carreau} can be used. In this model, the apparent viscosity $\mu$ is given by

\begin{equation}
\frac{\mu - \mu_{\infty}}{\mu_o - \mu_{\infty}}= \left(1+(\lambda\dot{\gamma})^a\right)^{\frac{n-1}{a}}
\label{eqn:carreau}
\end{equation}

\noindent where $n$, $a$, and $\lambda$ are empirically-determined material coefficients, and $\mu_o$ and $\mu_\infty$ are zero-shear viscosity and the viscosity at infinite shear-rate. In terms of $\tau$, Eq.\ \ref{eqn:carreau} can be written as:

\begin{equation}
	\frac{\tau - \tau_{\infty}}{\tau_o - \tau_{\infty}}= \left(1+(\lambda\dot{\gamma})^a\right)^{\frac{n-1}{a}}
	\label{eqn:carreauTau}
\end{equation}





	

\section{Foaming Model}

\subsection{Advection-Diffusion}

The advection and diffusion of the dissolved gas $c$ within the liquid and into bubbles is governed by an advection-diffusion equation:

\begin{equation}
   \frac{\partial c}{\partial t} + \nabla \cdot (c\mathbf{u}) = \nabla \cdot (D \nabla c)+ q
\end{equation}

\noindent where $q$ is a source term that takes into account gas generation due to, for example, chemical reactions. This equation is solved using another distribution function $g_i(\vx,t)$, such that the summation of $g_i$ gives the gas concentration (in terms of mass fraction) at location $\vx$ at time $t$:

\begin{equation}
c (\vx,t) = \sum_i^N g_i(\vx,t)
\end{equation}

\noindent The lattice Boltzmann equation for the evolution of $g_i(\vx,t)$ is given by:

\begin{equation}
\
g_i(\vx + \ei, t + 1) = g_i(\vx,t) + \frac{1}{\tau_g}\left(g_i^{eq}(\vx,t)-g_i(\vx,t)\right) + w_i q
\label{eqn:diffusion}
\end{equation}

\noindent We solve the advection-diffusion equation using lattices with fewer velocity vectors \cite{advectiondiffusion}: on D2Q5 (in 2D) or D3Q7 (in 3D) lattice topologies, that coincide with the fluid flow lattices. This calculation requires less computational time because of the reduced number of velocities. $\tau_g$ is the relaxation time for the advection-diffusion equation, which relates to the diffusion constant $D$ of the gas in the liquid according to the following equation:

\begin{equation}
D = c_s^2\left(\tau_g - \frac12 \right)
\label{eqn:diffusionconst}
\end{equation}

\noindent where $c_s^2$ is equal to $1/3$ and  $1/4$ for the D2Q5 and D3Q7 lattices, respectively.  The equilibrium distribution in the advection-diffusion equation is given by:

\begin{equation}
g_i^{eq}(\vx,t) = \wi c (\vx,t) \left(1 + \frac{\ei^\top \vec{u}}{c^2_s} \right)
\end{equation}

\noindent where $\vec{u}$ is calculated from Eq.\ \ref{eqn:velocity}.

The amount of gas diffusing into bubble $i$, $\Delta m_g^j$, is calculated based on the concentration gradient at the liquid-bubble interface:

\begin{equation}
	\normalsize
\Delta m_g^j = \rho \sum_{\vx \in I^j} \left(\sum_{\vx + \ei \in F}[g_{\bar{i}} (\vx + \ei,t)-g_i(\vx,t)] - c(\vx,t)[\alpha(\vx,t)-\alpha(\vx,t-1)]\right)
	\label{eqn:gascurrent}
\end{equation} 

\noindent where $I^j$ is the liquid-gas interface of bubble $j$. The last term in Eq.\ \ref{eqn:gascurrent} deducts the pure advection portion of gas transport, which is not diffused into the bubble.



\subsection{Henry's Law Boundary Condition}
\label{sec:HenrysLaw}
In many foam media such as polymers, the concentration of gas at a bubble interface obeys Henry's law, which expresses a linear relationship between pressure and gas concentration:

\begin{equation}
c(\vx,t)=k_Hp_g \qquad \forall \vx \in I
\label{eqn:henryslaw}
\end{equation} 

\noindent Similar to Eq.\ \ref{eqn:reconstruct}, this boundary condition is enforced by setting the unknown distribution coming from the gas cells $g^u_i$ to

\begin{equation}
g^u_i (\vx, t+1) = g^{eq}_i(c,\vec{u}_I) + g^{eq}_{\bar{i}}(c,\vec{u}_I) - g_{\bar{i}}(\vx, t + 1)
\label{eqn:reconstructdiffusion}
\end{equation}

\noindent where $c$ is given by Eq.\ \ref{eqn:henryslaw} \cite{Korner2005}.

\subsection{Disjoining Pressure}

To stabilize the lamella between bubbles, a disjoining pressure  $\Pi$ is assumed to be active within a maximum distance $d_{max}$ between two different bubble interfaces, with a linear dependence on the distance between the interface cells $d$ \cite{Korner2005}:

\begin{equation}
    \Pi=\begin{cases}
    0 & d>d_{max}\\
    k_{\Pi}(d_{max}-d) & d<d_{max}\\
    
    \end{cases}
\end{equation}

\noindent where $k_\Pi$ is a constant. In this work, the disjoining pressure is active up to four lattice cells i.e., $d_{max}=4\delta_l$.

\noindent This disjoining pressure is a repulsive force that resists bubble coalescence and can stabilize a lamellae. Physically, $\Pi$ originates from the variation of Gibbs free energy with the distance $d$ between two interfaces that correspond to different bubbles \cite{YUE2005163,Huber2014}. Similar to the surface tension implementation, the disjoining pressure is added to $\rho_g$:

\begin{equation}
\rho_g = \frac{p_g - 2\gamma\kappa(\vx) - \Pi}{c^2_s}\qquad \vx \in I
\label{eqn:rhobplusdisjoining}
\end{equation}

\begin{figure}[H]
	\centering

    \includegraphics[width=0.9\textwidth]{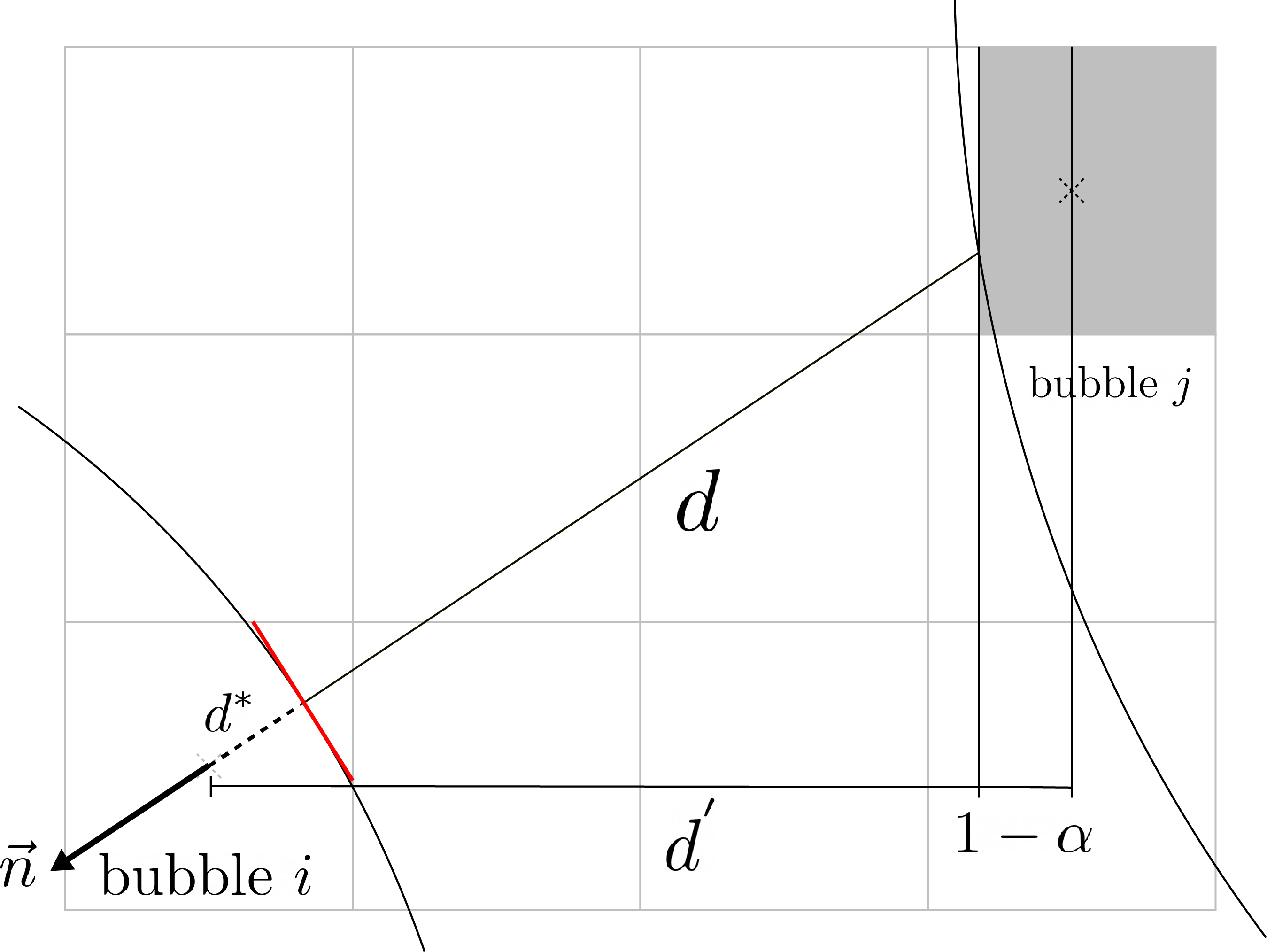}
	\caption{Calculating the distance between two interfaces belonging to adjacent bubbles. The reconstructed interface is shown in red.}
	\label{fig:disjoiningpressure}
\end{figure}

Calculating $\Pi$ is a two-step process as shown in Fig.\ \ref{fig:disjoiningpressure}. First, for each interface cell, a traversal ray tracing algorithm \cite{Amanatides} is used to move along the bubble $i$ interface normal $\vec{n}$ as far as $d_{max}$, in search of an interface belonging to another bubble $j$. If an interface cell is detected, $d^{'}$ is calculated as the width of the traversed cells minus $1-\alpha$, and projecting that distance along the normal vector to find $d+d^*$. The distance $d$ is then determined by subtracting the distance from the center of the interface cell to the interface $d^*$, which is calculated by geometrically reconstructing the interface of the bubble using a piecewise linear interface construction (PLIC) \cite{koerner_2008}.

A full description of the implementation of the ray tracing algorithm is given in \ref{apx:Alg}. Since we use cubic or square lattices, we use the explicit analytical expressions developed by Scardovelli and Zaleski \cite{SCARDOVELLI2000228} for PLIC calculations based on the implementations provided in Ref.\ \cite{VOFTOOLS}, which are faster than iterative methods.

\begin{figure}[H]
    \centering
    \includegraphics[width=0.9\textwidth]{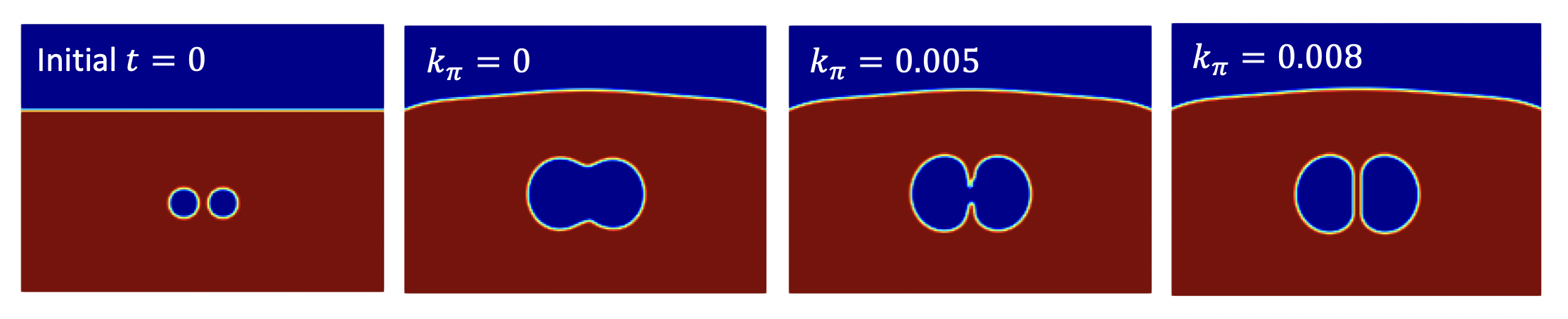}
    \caption{Effect of $\Pi$ on bubble coalescence. The leftmost image shows the initial condition at $t=0$ while the rest results are at $t=3400$.}
    \label{fig:kpieffect}
\end{figure}

Fig.\ \ref{fig:kpieffect} shows the effect of $\Pi$ on lamella stabilization between two growing bubbles.  In all the three cases, the initial parameters are identical, except for the value of $k_\Pi$ that regulates the strength of $\Pi$. The growing of the bubbles draw the bubbles together until at $t=2800$, in the absence of a disjoining pressure ($k_\Pi=0$), the bubbles coalesce upon contact (the exact timestamp is not shown in Fig.\ \ref{fig:kpieffect}), and no stable lamella forms. At $k_\Pi=0.005$ other forces overcome the disjoining pressure and the bubbles still coalesce, but the disjoining pressure delays the coalescence until $t=3400$. At $k_\Pi=0.08$, the disjoining pressure is strong enough to prevent coalescence, and so a stable lamella forms between the bubbles.

\begin{figure}[H]
    \centering
    \includegraphics[width=\textwidth]{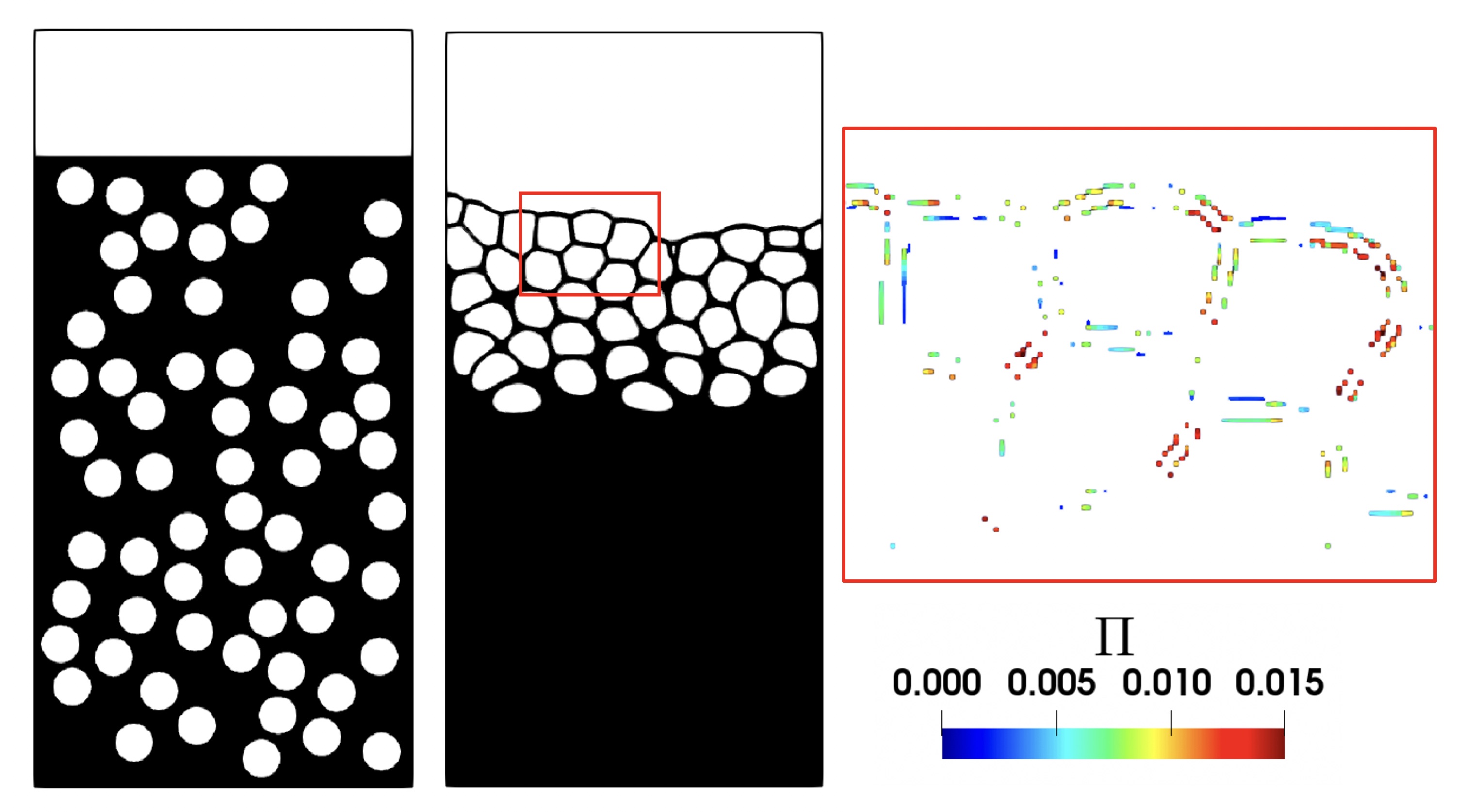}
    \caption{Bubbles ascending due to the effect of gravity. Left: Initial configuration of bubbles. Middle: Bubbles ascend due to buoyancy. Right: The disjoining pressure field shown for a selected portion of the middle image shown with the red box. $k_{\Pi} = 0.005$ and $d_{max}=4\delta_l$. }
    \label{fig:latte}
\end{figure}

\noindent The formation of lamella is more apparent in a system with multiple bubbles. Fig.\ \ref{fig:latte} shows a number of bubbles ascending in a bubble column due to buoyancy. Although 18 bubbles coalesce, the disjoining pressure eventually stabilizes the lamella between the remaining bubbles and enables the formation of a cellular structure near the liquid surface.

\subsection{Nucleation}

\begin{figure}[H]
    \centering
    \includegraphics[width=\textwidth]{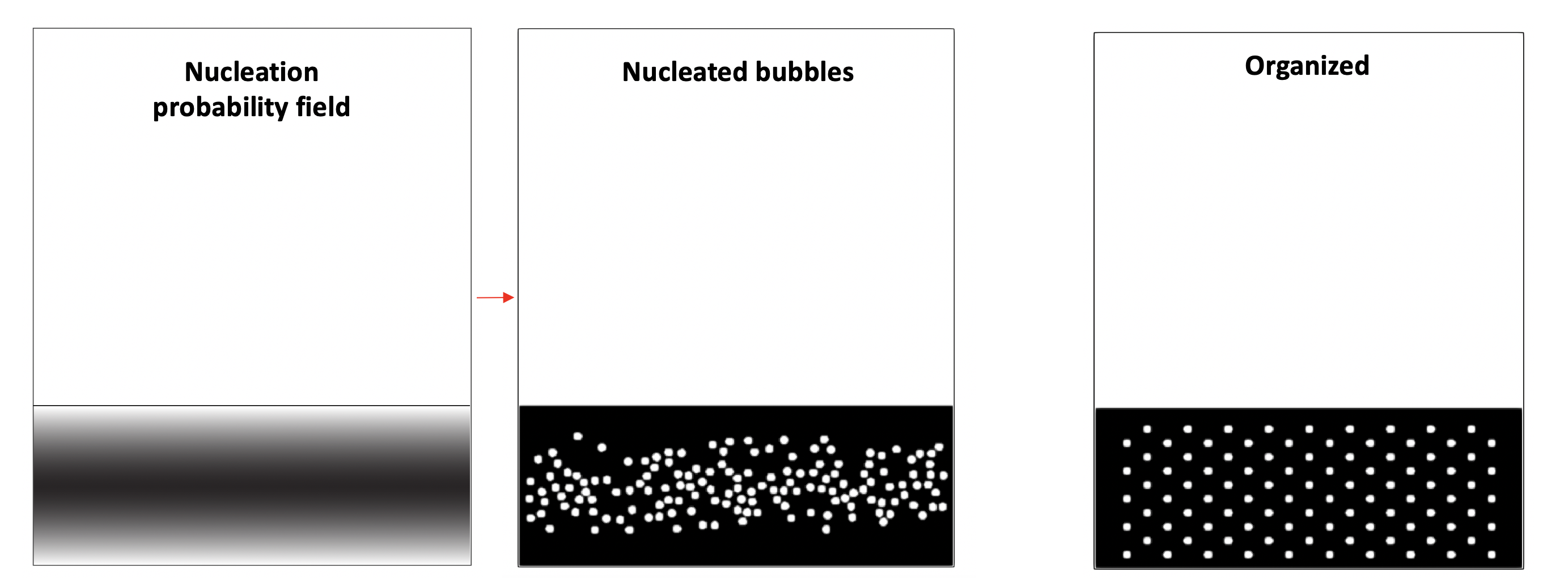}
    \caption{Given a nucleation probability field, a disk sampling algorithm distributes the bubbles subject to a minimum distance (left). Bubbles can be distributed in an organized fashion (right).}
    \label{fig:nucleation}
\end{figure}

The location and radius of nucleated bubbles must be specified at the beginning of each simulation. A nucleation probability field (derived on the basis of a nucleation model) may be used to specify the distribution and initial radius of the nucleated bubbles, as shown in Fig.\ \ref{fig:nucleation}. If we pick random points in the domain, the nucleated bubbles may overlap, as the bubbles are disks (2D) or spheres (3D), not points. For this purpose, a Poisson Disk Sampling algorithm is implemented based on the work of \cite{disksampling}, which enforces that the distance between selected nucleation sites be at least equal to the radius of the bubbles plus one $\delta_l$. Bubbles can also be distributed in an ``organized'' fashion (see Fig.\ \ref{fig:nucleation}). 

The initial configuration of the bubbles have a major effect on the final structure of the foam. Section \ref{sec:bubbleDisEffect} presents an example of the effect of different initial bubble distributions.

\subsection{Software Structure}

\begin{figure}
    \centering
    \includegraphics[width=0.25\textwidth]{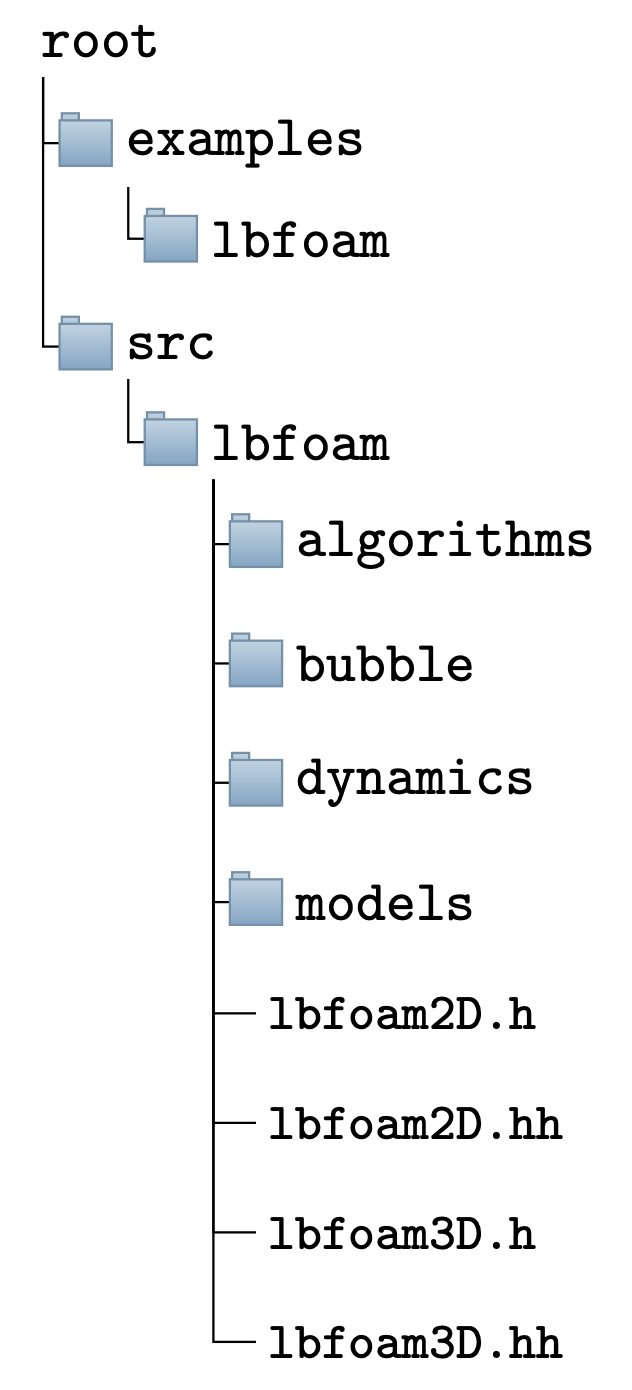}
    \caption{LBfoam directory structure.}
    \label{filetree}
\end{figure}

Fig.\ \ref{filetree} shows the directory structure of the LBfoam library. Many of the LBfoam functions and classes either call or extend the Palabos library classes. The Palabos open source framework provides basic tools for parallelization of LBM simulations using Message Passing Interface (MPI) and Lattice Boltmzann solvers. LBfoam is written in a non-intrusive manner with respect to Palabos, except for a few exceptions, to allow researchers to use it along with future versions of Palabos. Currently LBfoam uses Palabos v2.0, which is the latest version.

\noindent The algorithms folder contains the PLIC, ray-tracing, Poisson disk-sampling algorithm, and other related functions. The $dynamics$ folder contains the classes for advection-diffusion coupling and applying the Henry's law boundary condition. The $bubble$ folder includes the classes for calculating gas diffusion into each bubble, updating bubble pressures, tracking bubble coalescence, calculation of disjoining pressure, and updating bubbles gas content. The $models$ folder includes a 2D FSLBM model which is not available in Palabos. The directory $examples/lbfoam$ include a variety of demo cases. With the ongoing development of the library, this directory structure may change in the future.

$lbfoam2D.h$ and $lbfoam3D.h$ header files provide access to LBfoam classes for 2D and 3D simulations, respectively. Each LBfoam function/class must be called using the namespace called {\tt lbfoam} (e.g.\ {\tt lbfoam::PLIC2D}).

Palabos uses a data-structure called ``multi-block" that is composed of a number of Cartesian meshes, that combined create a computational domain. These cartesian meshes are distributed to a number of processing units. At each iteration, every processor tags the cells  on its sub-domain that belong to a bubble with a unique ID using a flood-fill algorithm, and attaches bubble information such as volume and gas content to the ID. The ID of each bubble is unique and global among processors. In the case of foaming, there are two situations that require special communication (data transfer) between processors. The first relates to the calculation of the disjoining pressure: when a bubble interface is closer than $d_{max}$ from a sub-domain boundary, an adjacent bubble could be in another sub-domain. In this case, the communication envelope between the two processors must be expanded as much as $d_{max}$. Second, when a bubble occupies multiple sub-domains, bubble gas content must be shared among all sub-domain processors. Since gas can diffuse into a bubble from different sub-domains, this communication is necessary to calculate the correct value of gas content at each iteration. Bubble merging or splitting is carefully monitored at each iteration as the gas content must be distributed to the resulting bubble(s) in proportion to their volume(s).

While bubble tracking involves multiple communications between distributed CPUs, we evaluated the performance of our model on up to 100 nodes (with 40 processors each) on the SciNet Niagara supercomputer \cite{scinet} at the University of Toronto, and were able to achieve close to linear parallel scaling.

\section{Validation and Sample results}

\subsection{Advection-Diffusion}

\begin{figure}[H]
	\centering
	\includegraphics[width=0.7\textwidth]{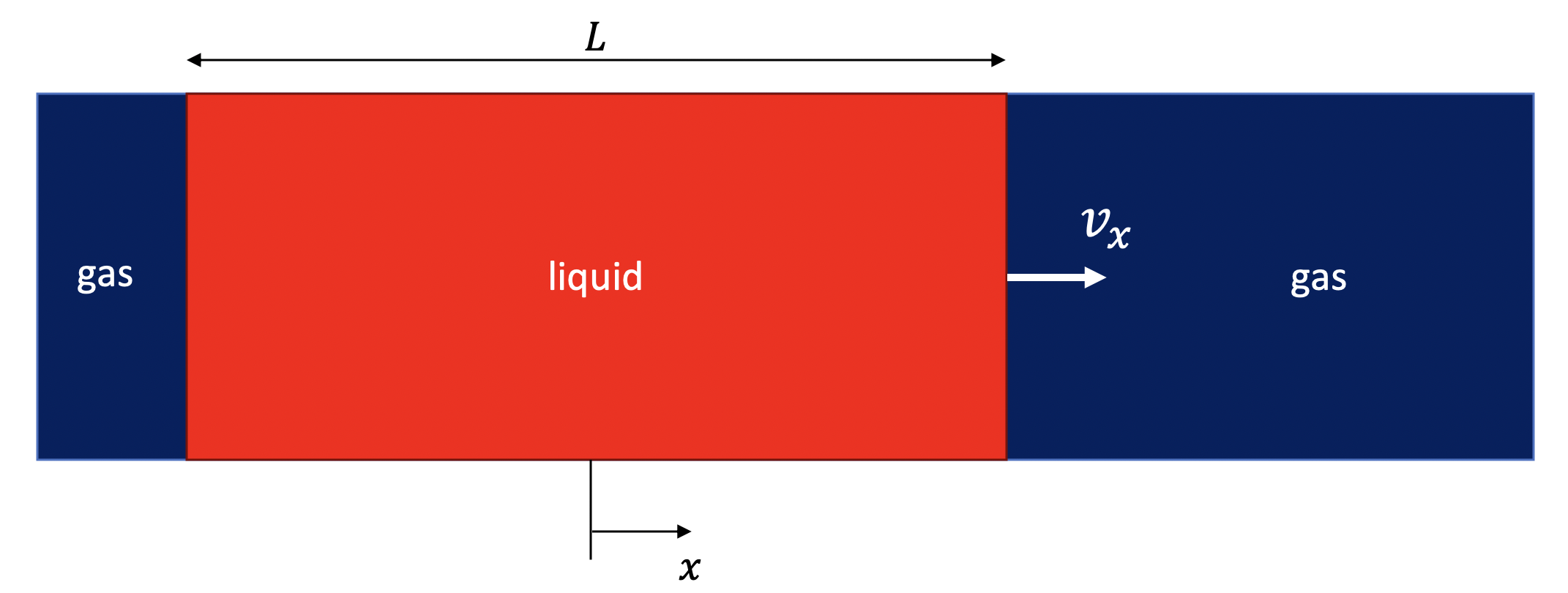}
	\caption{Schematic of the advection-diffusion problem. }
	\label{fig:diffusionValidation}
\end{figure}

We begin by examining the solution of the advection-diffusion equation to an available analytical solution. The schematic of the problem is shown in Fig.\ \ref{fig:diffusionValidation}. A 1D section of liquid (in red) of length $L$ moves at a constant velocity ($v_x$) within a rectangular domain, and is surrounded by gas on both sides. Boundaries are periodic, and gas is generated at a constant rate within the liquid ($q$ is the source term), and diffuses into the gas phase across the interfaces. The steady-state analytical solution for the concentration $c$ is given by

\begin{equation}
    \frac{8D(c-k_{H}p_g)}{q L^2} = 1-x'
    \label{eqn:diffusionanalytical}
\end{equation}

\noindent where $x'=2x/L$. 

\begin{figure}[H]
	\centering
	\includegraphics[width=0.7\textwidth]{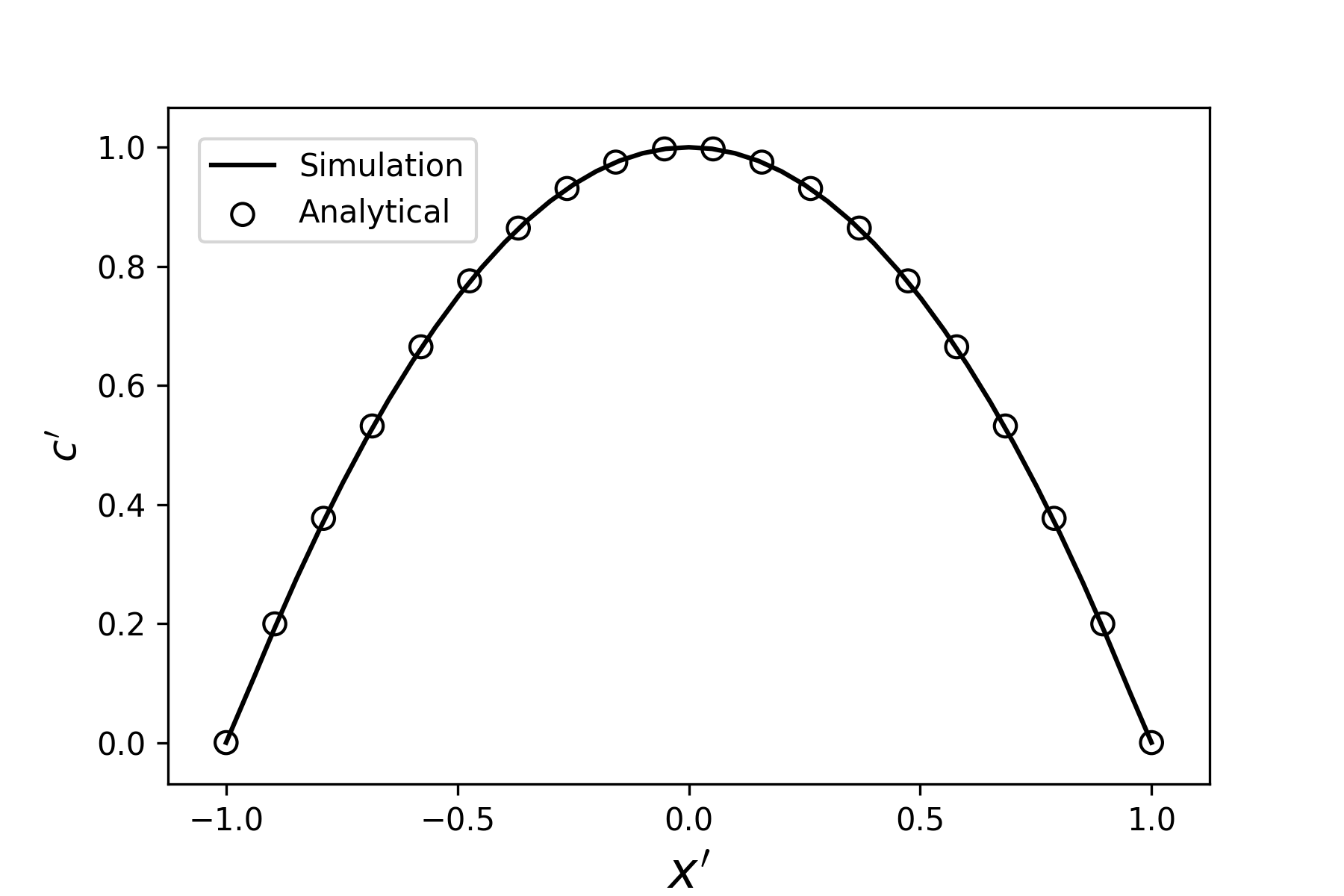}
	\caption{A comparison between an LBfoam result and an analytical steady-state result for the system shown in Fig.\ \ref{fig:diffusionValidation}}
	\label{fig:CvsX}
\end{figure}

\setlength{\tabcolsep}{20pt}
\renewcommand{\arraystretch}{1.5}
\begin{table*}[t]
	\centering
	\caption{Advection-diffusion simulation parameters.}
	\begin{tabular}{lccc}
		\specialrule{.15em}{.05em}{.05em} 
		
            $v_x$ & $0.2$ \\ 
            $q$ & $1e^{-6}$\\ 
            $\nu$  & $0.25$\\ 
            $D$ & $0.06$\\ 
            domain size & $200 \times 20$  \\
            $k_H$ & $0.01$\\
            $p_g$ & $\frac13$\\
            $L$  & $40$\\ 
		              
		 \specialrule{.15em}{.05em}{.05em} 

	\end{tabular}
	\label{table:validatonParameters}
	
\end{table*}

\noindent The simulation parameters are given in Table \ref{table:validatonParameters}. Fig.\ \ref{fig:CvsX} shows that the simulation result matches the analytical solution, demonstrating that the advection-diffusion equation is being correctly solved, and that Henry's law on the moving interface boundaries is being properly applied. We confirmed this result using other parameter sets, and successfully conducted the grid independence test using domain sizes with $0.5$ and $2x$ resolution of the domain size presented here.

\subsection{3D Bubble Growth}

\noindent The following is the analytical solution to the 1D ``Cell Model" for a single bubble growing in an infinite reservoir (i.e.\ the gas concentration at infinity $c_{\infty}$ is constant \cite{SCRIVEN19591}):

\begin{equation}
    R(t) = \sqrt{2 \Delta c V_m Dt + R_0^2}
    \label{eqn:1Danalytical}
\end{equation}

\noindent $R(t)$ is bubble radius at time $t$, $R_0$ is the initial bubble radius, $V_m = RT/p$ is the molar volume and $\Delta c = c_{\infty} - c_R$ where $c_R = k_Hp_g$.  

\begin{figure}[H]
	\centering
	\includegraphics[width=0.6\textwidth]{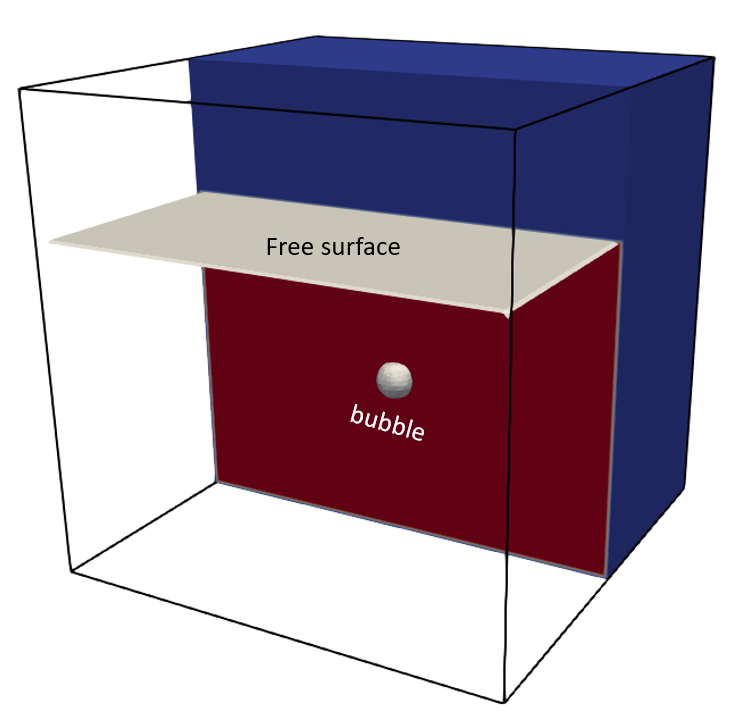}\
	\caption{Initial configuration of a 3D LBM simulation of single bubble growth. The domain consists of $100\times100\times 100$ cells.}
	\label{fig:advdiffinitial}
\end{figure}

\begin{figure}[H]
	\centering
	\includegraphics[width=0.8\linewidth]{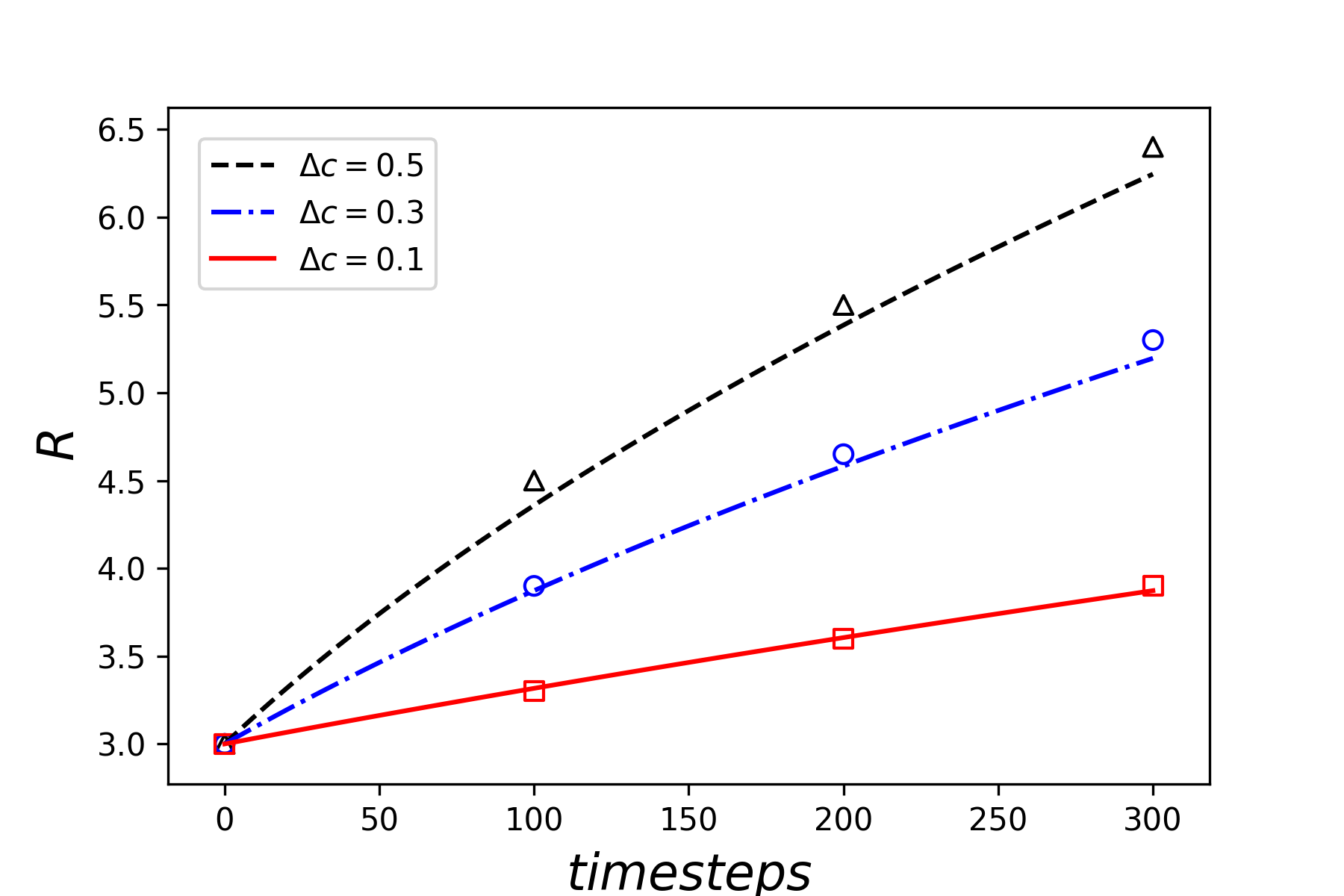}
\caption{Comparison of LBfoam results and an analytical bubble growth model for the growth of a single bubble in a large reservoir. Square, circular, and triangular markers correspond to LBfoam results for $\Delta c = 0.1, 0.3$ and $0.5$, respectively.}
	\label{fig:comparisonRvsT}

\end{figure}

\noindent To solve the same problem in LBfoam, a bubble nucleus is placed in a reservoir 30 times larger than the initial radius of the bubble, so that for the duration of the simulation $c_{\infty}$ remains almost constant (see Fig.\ \ref{fig:advdiffinitial}). In the simulation, $R_0=3\delta_l$, $k_H=0.001$, $D=0.03$, $V_m=3$, $p_0=1/3$, $\rho=1$, $\gamma=g=0$, and $\nu=0.25$, and we varied $c_{\infty}$ so that $\Delta c = 0.1, 0.3$ and $0.5$. The radius of bubble was calculated as $R = \sqrt[3]{3V_b/4 \pi}$ where $V_b$ is the bubble volume (although the bubble remained almost spherical throughout the simulations). The comparison between Eq.\ {\ref{eqn:1Danalytical}} and LBfoam is given in Fig.\ \ref{fig:comparisonRvsT} and shows very good agreement. The small discrepancy between the results can be attributed to the assumption that $c_{\infty}$ remains constant throughout the simulation, and the related effect of the domain boundaries on the growth of the bubble.

\subsection{Bubble Dissolution}

In foam injection molding (FIM), bubbles nucleate due to a pressure drop at the inlet gate during filling. Due to large shear stresses during the mold-filling process, these gate-nucleated bubbles become undesirably elongated  \cite{Shaayegan2016}. To promote foam uniformity, in high-pressure FIM, these bubbles are re-dissolved into the polymer by increasing the cavity pressure to promote the solubility of the gas, where the gas diffuses from the bubbles back into the polymer melt. Uniform bubbles are then nucleated again by a second pressure drop (due to melt shrinkage or mold opening) \cite{TROMM201943}. 

\begin{figure}[H]
	\centering
	\includegraphics[width=\textwidth]{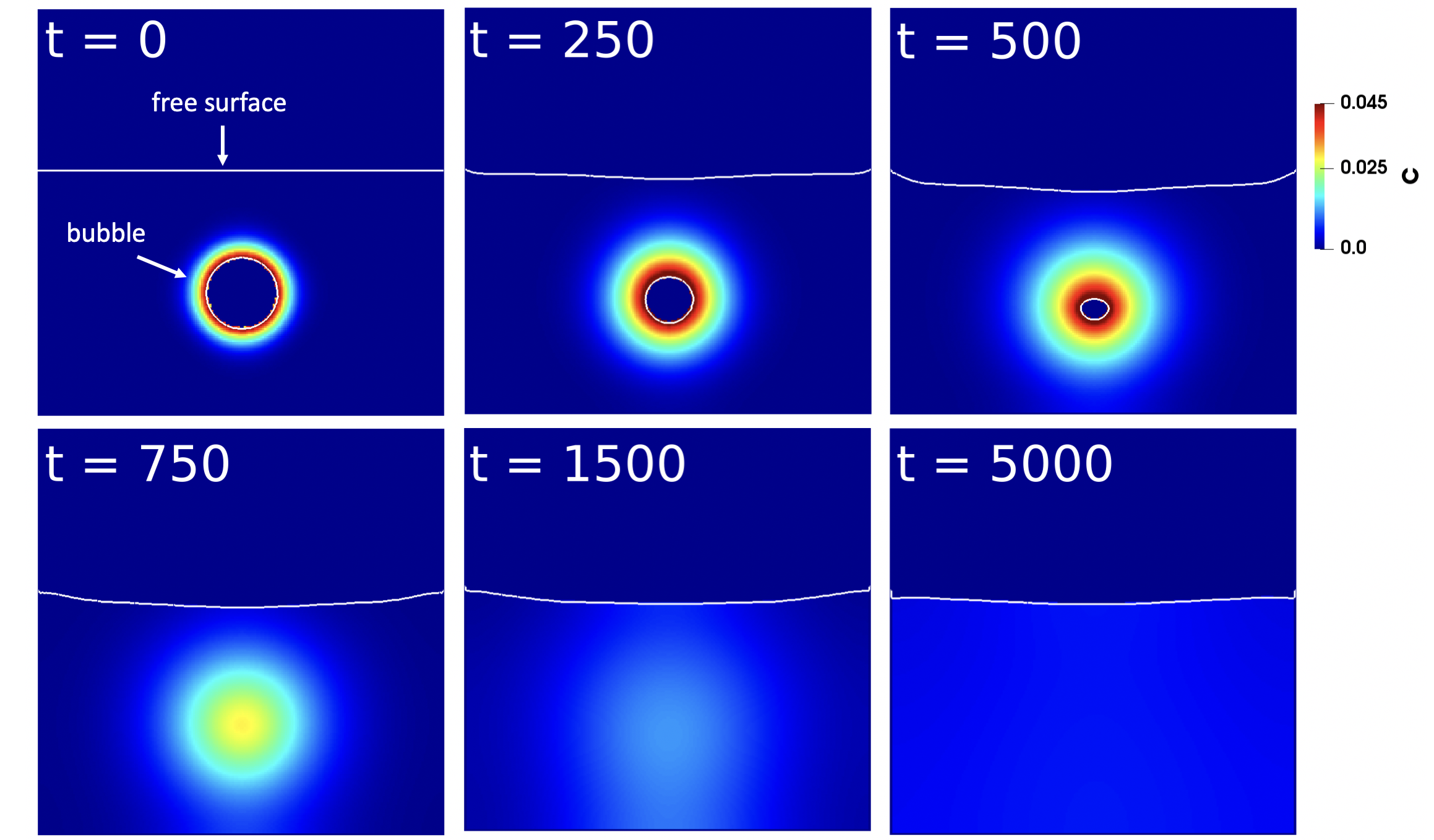}%
	\caption{Dissolution of a bubble in a liquid with no initial gas content. The concentration gradient around the bubble at $t=0$ is due to the Henry's law boundary condition at the liquid-bubble interface. Note that the final free surface height is lower at $t=5000$ compared to $t=0$\ due to the conservation of mass of the liquid. \textit{Parameters:} $\tau = 0.98 $, $\tau_g = 0.8$, $RT = 500$, $c_{0} = 0$, $k_{\Pi} = 0 $, $\rho = 1$, $k_H = 5\times 10^{-3}$, $g = 0$, $\gamma = 4.2\times 10^{-3}$.}
	\label{fig:dissolution}
\end{figure}

\noindent LBfoam can be used to study bubble dissolution. In Fig.\ \ref{fig:dissolution}, a bubble is placed in the middle of a liquid with no initial gas content, and with the source term set to zero. The non-zero concentration of gas at the bubble-liquid interface results in the diffusion of gas from the bubble to the liquid, until the bubble is fully dissolved.

\subsection{Foaming}

In the following, we parameterize simulations based on lattice units. The conversion between lattice units and SI units is straightforward and can be done by defining a length scale and a reference density for the simulation. Examples of unit conversions for foaming simulations are given in \cite{Korner2002,Korner2005}. The system parameters for each simulation are given in the figure captions. 

\begin{figure}[H]
    \centering
    \includegraphics[width=\textwidth]{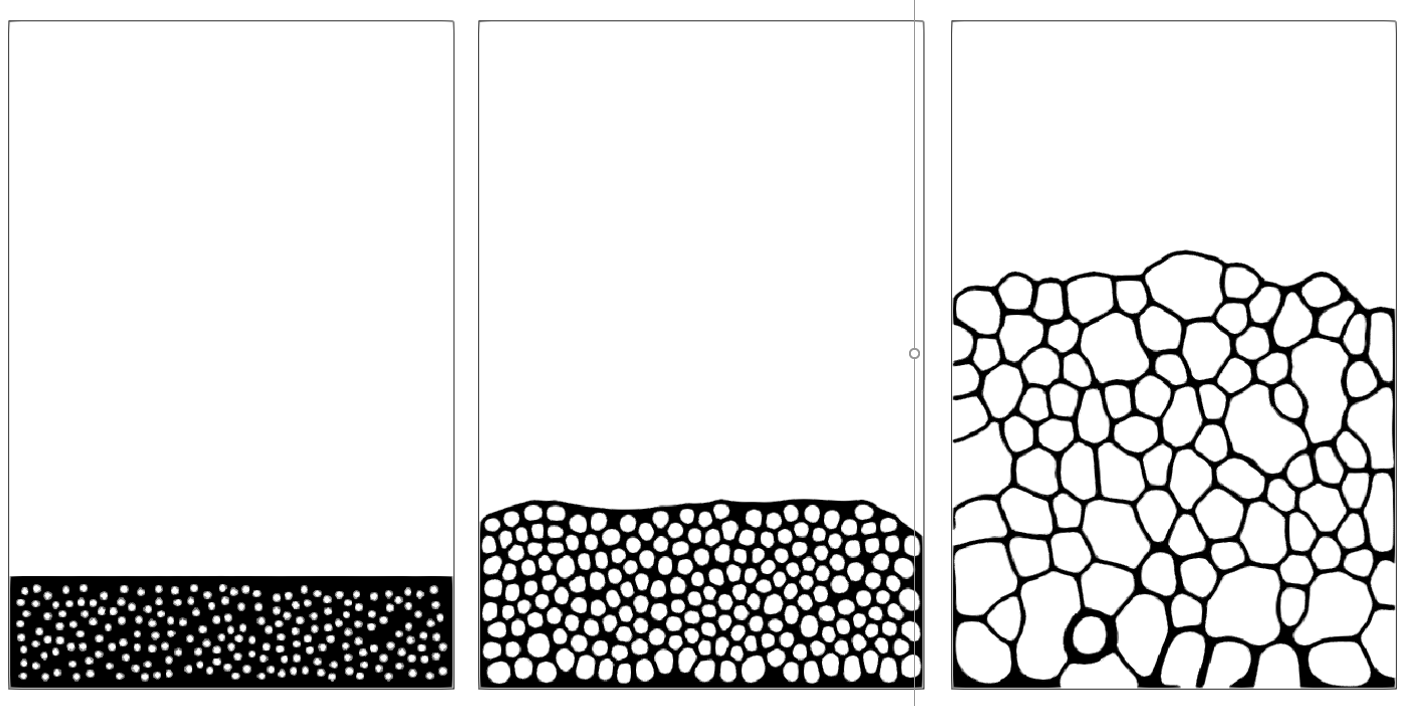}
    \caption{2D foaming \textit{Parameters:} $\tau = 0.9 $, $\tau_g = 0.6$, $RT = 1$, $c_{0} = 0.0$, $k_{\Pi} = 2\times10^{-3} $, $\rho = 1$, $k_H = 10^{-3}$, $g = 0$, $\gamma = 5\times 10^{-3}$, $q=5\times 10^{-5}$ $num.\ nuclei = 300$.}
    \label{fig:2Dfoam}
\end{figure}

2D foams have been investigated in fields such as rheology and microfluidics and many others (e.g. \cite{Marchalot_2008}). LBfoam is capable of simulating 2D foams as shown in Fig.\ \ref{fig:2Dfoam}, in a rectangular container initially containing 300 nuclei. The simulation domain consists of $450\times650$ cells. Initially a portion of the domain is filled with the liquid medium, gravity $g=0$ and, the nuclei are randomly distributed, each with a radius of $3$ cells. The initial gas concentration $c_0$ is zero; instead the gas is generated by a source term. Note that some of the bubbles coalesce with other bubbles during the simulation.

\begin{figure}[H]
	\centering
	\includegraphics[width=\textwidth]{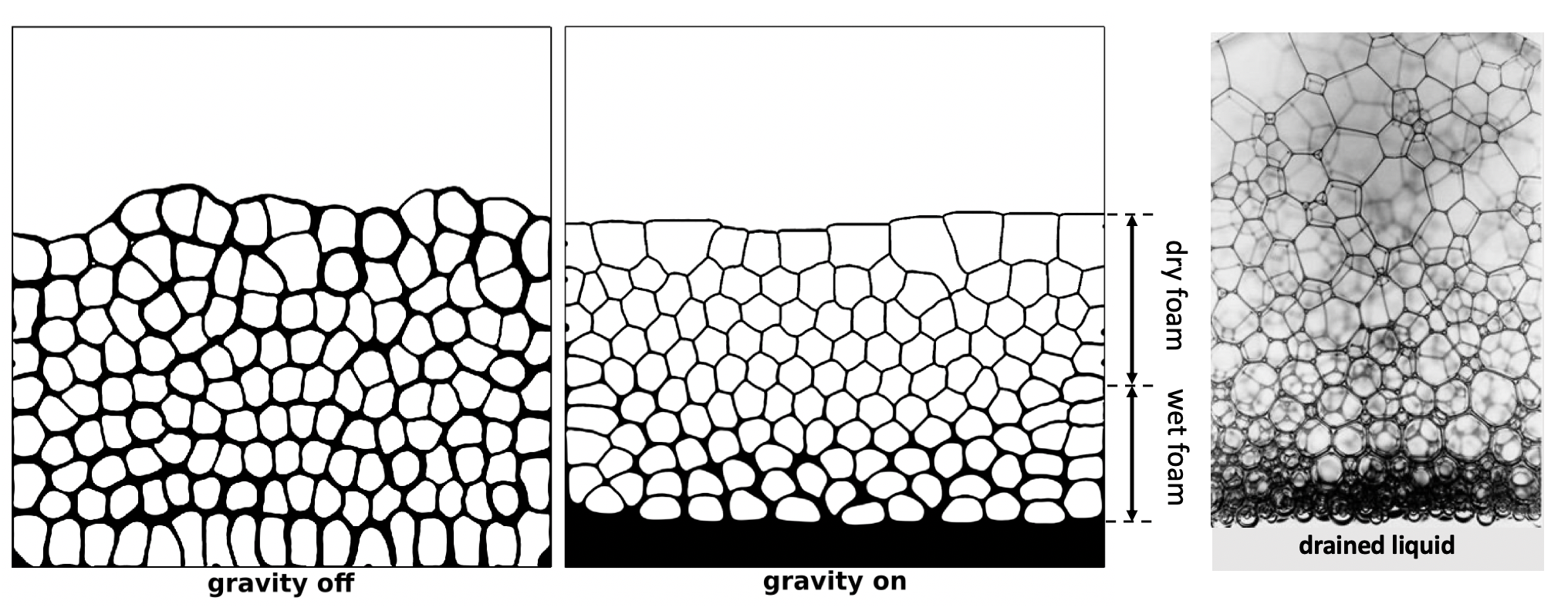}%
	\caption{The influence of gravity on foam drainage. In both simulations $t=30000$. \textbf{Left:} \textit{Parameters:} $\tau = 0.95 $, $\tau_g = 0.53$, $RT = 50$, $c_{0} = 0.02$, $k_{\Pi} = 2\times10^{-3} $, $q=0$, $\rho = 1$, $k_H = 10^{-5}$, $g = 0$, $\gamma = 5\times 10^{-3}$, $num.\ bubbles = 150$, $distribution: uniform$. \textbf{Middle:} \textit{Parameters:} Same as (a) except $g=3\times 10^{-5}$. \textbf{Right:} Foam drainage experimental result courtesy of European Space Agency.}
	\label{fig:foamDrainage}
\end{figure}

Under the influence of gravity the behavior is different, as liquid gradually moves to the bottom of the container, leaving the top of the foam dry of liquid. This effect is known as foam drainage, and has been extensively studied for Newtonian and non-Newtonian fluids (e.g.\ \cite{DRAINAGE1,DRAINAGE2,DRAINAGE3,DRAINAGE4,DRAINAGE5,DRAINAGE6}), previous foam drainage studies have been largely experimental; little mathematical and numerical modeling has been done. As shown in Fig.\ \ref{fig:foamDrainage}, foam drainage can be observed in the simulation results, in the presence of gravity. The liquid accumulates at the bottom of the container. It can also be seen that after drainage large sections of the foam evolve into honeycomb structures forming dry and wet foam sections (see Fig.\ \ref{fig:foamDrainage}) similar to experimental observations \cite{wetdryfoam}. The natural emergence of foam drainage illustrates the capability of the model for foam drainage studies of complex foaming phenomena.

\begin{figure}[H]
	\centering
	\includegraphics[width=0.8\textwidth]{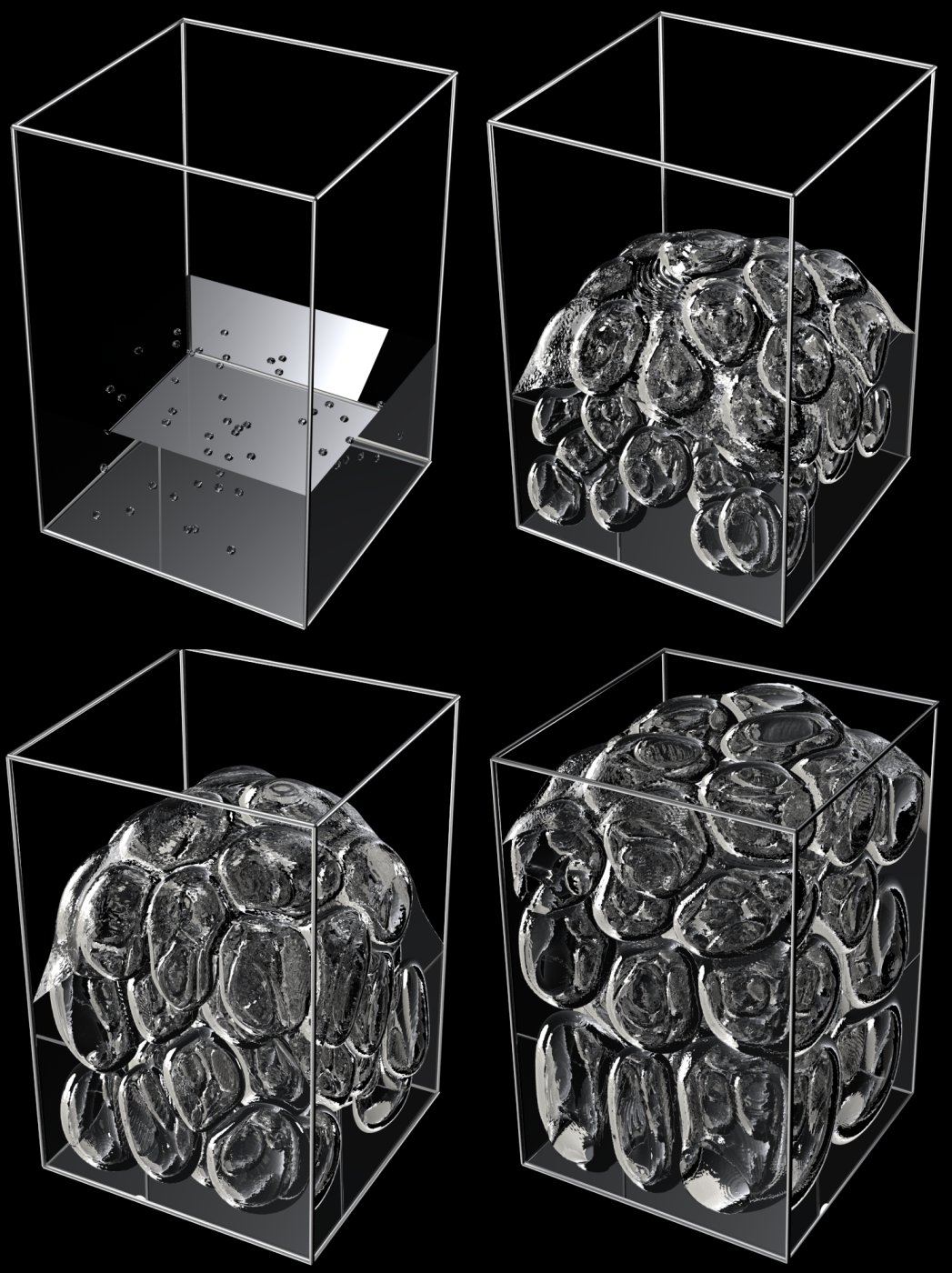}
	\caption{3D simulation of foaming with 40 randomly distributed nuclei in a domain of $300\times200\times200$ cells. The time steps from top left to bottom right are 0, 5000, 10000, and 30000, respectively.  \textit{Parameters:} $\tau = 0.95 $, $\tau_g = 0.53$, $RT = 50$, $c_{0} = 0.018$, $k_{\Pi} = 5\times10^{-3} $, $q=0$, $\rho = 1$, $k_H = 10^{-5}$, $g = 0$, $\gamma = 5\times 10^{-3}$, $num.\ bubbles = 40$.}
	\label{fig:3D}
	
\end{figure}

An example of a 3D simulation is given in Fig.\ \ref{fig:3D}. 40 bubble nuclei are randomly distributed in a liquid medium, each with an initial radius of 3 cells. The domain is made up of $300\times200\times200$ cells, or 12 million in total.

\begin{figure}[H]
    \centering
    \includegraphics[width=\textwidth]{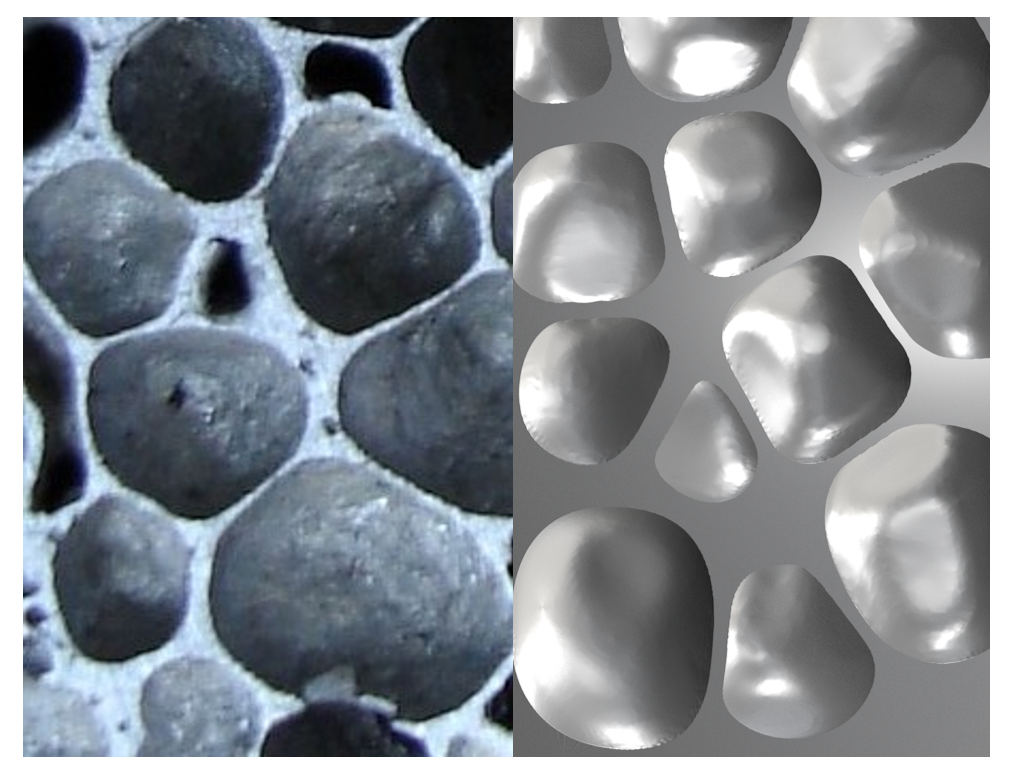}
    \caption{A 3D LBfoam result (right) resemblse the foam structure in an aluminum foam (image from Wikimedia/Creative Commons)}
    \label{fig:3Dfoam}
\end{figure}

\noindent In Fig.\ \ref{fig:3Dfoam} we illustrate a comparison between a cross-section of the 3D simulation result and an aluminium foam, that shows good qualitative agreement.

\subsection{The effect of Nuclei Distribution}
\label{sec:bubbleDisEffect}

\begin{figure}[H]
    \centering
    \includegraphics[width=0.6\textwidth]{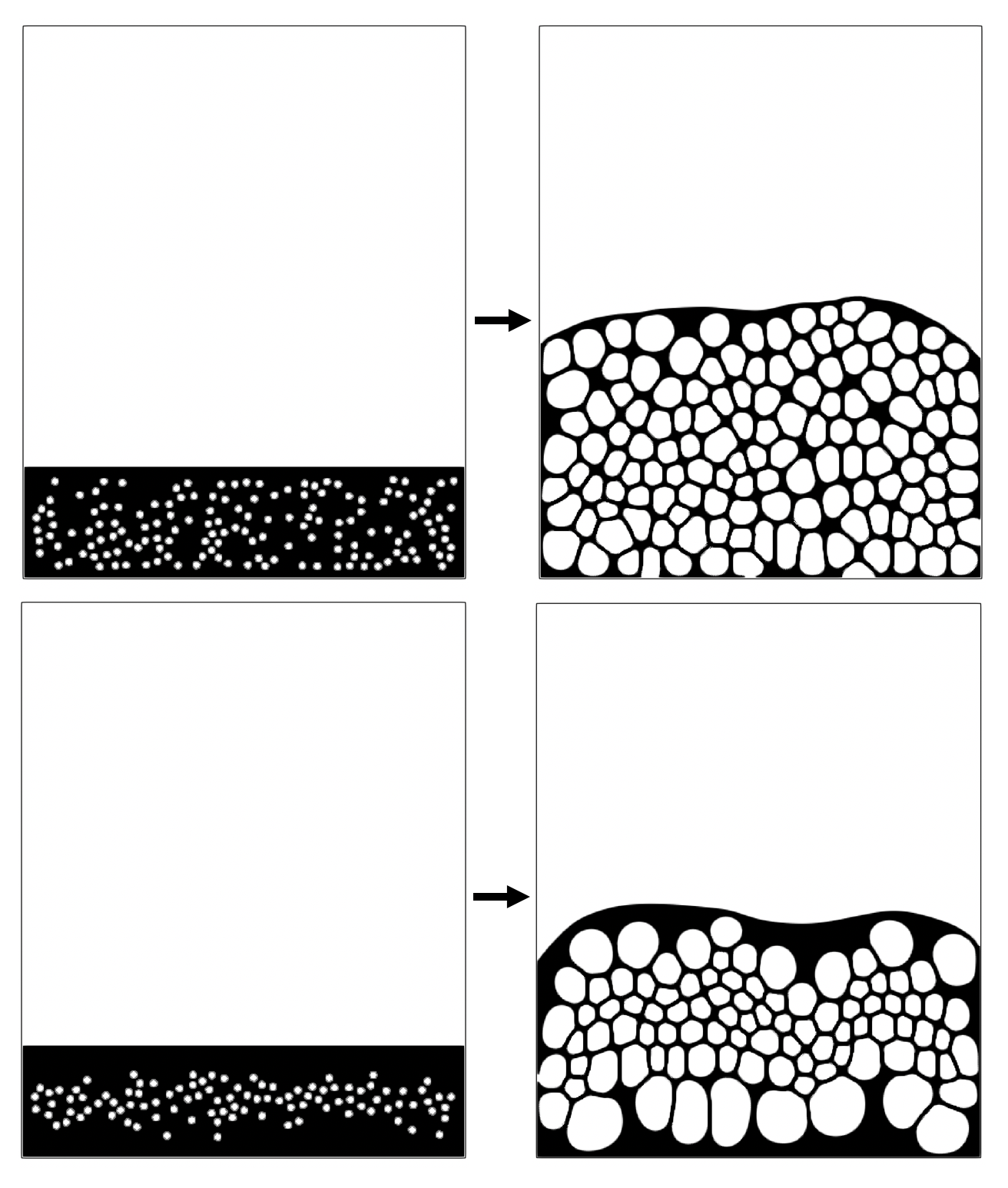}
    \caption{Effect of initial bubble nuclei distribution on the foam structure. Left: $t=0$, right: $t=13600\delta_t$ \textit{Parameters:} $400\times 500$ domain, $\tau = 0.9$, $\tau_g = 0.6$, $RT = 1$, $c_{0} = 0$, $k_{\Pi} = 0.008$, $\rho = 1$, $k_H = 5\times 10^{-3}$, $g = 0$, $\gamma = 4.2\times 10^{-3}$.} 
    \label{fig:bubbleDisEffect}
\end{figure}

The initial nuclei distribution can affect the foam structure. Fig.\ \ref{fig:bubbleDisEffect} demonstrates two cases, one in which the bubbles are initially distributed evenly, and one in which the bubbles are nucleated near the center of the liquid. This can happen in conditions where, for example, the liquid center is hotter and promotes nucleation. As shown in Fig.\ \ref{fig:bubbleDisEffect}, the different initial distributions result in different structure and expansion of the foam.

\section{Conclusion}

We have developed a simulation software based on the lattice Boltzmann Method that can simulate foaming in 2D and 3D. An advection-diffusion equation is coupled to a fluid flow solver to account for dissolved gas diffusion into bubbles, subject to a Henry's law boundary condition at bubble-liquid interfaces. A volume-of-fluid method is used to track and locate bubble interfaces. The model accounts for disjoining pressure between bubbles, surface tension, film drainage, and bubble dynamics, including bubble growth, deformation, coalescence, bursting and splitting. To incorporate the disjoining pressure into the model, a fast traversal algorithm is used to calculate the distance between adjacent bubbles, and the bubble interfaces are reconstructed via a PLIC algorithm. The library is based on the Palabos library, which enables large scale parallel simulations. Simulation results demonstrate the potential of the model to provide a better understanding of the evolution, structure, and rheology of foams. 

The software is publicly available under the AGPL v3 license at the following Github repository, and is being continuously improved:

https://github.com/mehdiataei/LBfoam

\section*{Acknowledgements} \nonumber
We thank the Natural Sciences and Engineering Research Council of Canada (NSERC) and Autodesk Inc.\ for their financial support. Computations were performed on the Niagara supercomputer at the SciNet HPC Consortium. SciNet is funded by: the Canada Foundation for Innovation; the Government of Ontario; Ontario Research Fund - Research Excellence; and the University of Toronto.

%

\bibliographystyle{unsrtnat}


\bibliography{p}

\appendix

\newpage
\section{Ray Tracing Traversal Algorithm}
\label{apx:Alg}

The ray tracing algorithm for detection of adjacent bubbles is as follows:

For each interface cell, the algorithm marches along the interface normal $\vec{n}$ away from the bubble, until it finds an interface that belongs to another bubble. Referencing the traversal algorithm in 2D described in Algorithm\ \ref{alg:traversal}, for an interface cell $i$, the line along the negative direction of the interface normal is defined as:
\begin{align}
\underline{n}=-\vec{n}=
\beta\begin{bmatrix}
-n_{x} \\
-n_{y} \\
\end{bmatrix} \qquad \beta \geq0
\end{align}

\noindent where $n_x$ and $n_y$ are the components of the normal vector. First, the algorithm calculates the values of $\beta$ required for $\underline{n}$ to cross the vertical ($betaMaxX)$) and horizontal ($betaMaxY$) cell boundaries of cell $i$ ($betaMaxX$). The first neighboring cell along $\underline{n}$ is determined by the minimum of $betaMaxX$ and $betaMaxY$.  In order to locate the subsequent neighboring cells along $\underline{n}$, the minimum of $betaMaxX$ and $betaMaxY$ is found after incrementing by the value of $\beta$ required to move along $\underline{n}$ for one cell length. Extending the algorithm to 3D is trivial and only requires calculating $betaMaxZ$ in the $z$ direction to calculate the minimum of $betaMaxX$, $betaMaxY$ and $betaMaxZ$.

\begin{spacing}{0.7}	 
	\SetKwInOut{Parameters}{parameters}
	\SetKw{Continue}{continue}
	\SetKw{Break}{break}
	\SetKw{Return}{return}
	\begin{algorithm}
		\footnotesize
		\SetKwData{Left}{left}
		\SetKwData{This}{this}
		\SetKwData{Up}{up}
		\SetKwFunction{Union}{Union}
		\SetKwFunction{FindCompress}{FindCompress}
		\SetKwInOut{Input}{input}
		\SetKwInOut{Output}{output}
		\vspace{1em}
		$d_{max}$: maximum marching distance  \\
		$nx$, $ny$: outward normal components \\
		$x$, $y$: cell coordinates \\
		$I$: set of interface cells \\
		$BC$: set of boundary cells \\
		$ID$: bubble IDs \\
		
		\ForEach{$(nx,ny) \in I$ }{
			\tcp{ The value of $\beta$ required to exit the initial vertical and horizontal cell boundaries.}	 	
			
			$betaMaxX \leftarrow abs(0.5 / nx)$\;
			$betaMaxY \leftarrow abs(0.5 / ny)$\;
			\tcp{ The value of $\beta$ required to move the length of one vertical and horizontal cell boundary.}	 
			
			$betaDeltaX \leftarrow 2 * betaMaxX$\;
			$betaDeltaY \leftarrow 2 * betaMaxY$\;
			\tcp{$stepX$ and $stepY$ Determine direction of the marching algorithm.}	
			$stepX \leftarrow nx >= 0.0\ ?\ 1\ :\ -1$\;
			$stepY \leftarrow ny >= 0.0\ ?\ 1\ :\ -1$\;
			\tcp{$nextX$ and $nextY$ are initialized with the starting cell coordinates. }
			$nextX  \leftarrow x$\;
			$nextY  \leftarrow y$\;	
		}	
		
		\For{$i\leftarrow 1$ \KwTo $d_{max}$}{
			\uIf{$betaMaxX < betaMaxY$}{
				betaMaxX += betaDeltaX\;
				nextX += stepX\;
			}
			\uElseIf{$betaMaxX == betaMaxY$}{
				betaMaxX += betaDeltaX\;
				betaMaxY += betaDeltaY\;
				nextX += stepX\;
				nextY += stepY\;
			}
			\Else{
				betaMaxY += betaDeltaY\;
				nextY += stepY\;
			}
			
			\If{$ID(nx,ny) == ID(nextX,nextY)$ }{
				\Continue\;
				
			}
			\If{ $(nextX,nextY)$ $\in BC$ }{
				\Break\;	
			} 	\If {$(nextX,nextY)$ $\in I$} {
				\Return $(nextX,nextY)$\;
				
			}
			
		}
		
		\caption{Traversal algorithm}\label{algo_disjdecomp}
		\label{alg:traversal}
		
	\end{algorithm}

\end{spacing}

\end{document}